\documentclass[authoryear,preprint,review,12pt]{elsarticle}

\usepackage{amsmath}
\usepackage{amsfonts}
\usepackage{graphicx,floatrow}
\usepackage{subfigure}
\usepackage{epsfig}
\usepackage{natbib}
\usepackage{booktabs}
\usepackage{multirow}
\usepackage{float}

\usepackage[colorlinks=true,breaklinks=true,bookmarks=true,urlcolor=blue,
     citecolor=blue,linkcolor=blue,bookmarksopen=false,draft=false]{hyperref}

\newtheorem{theorem}{Theorem}

\newtheorem{definition}{Definition}

\newtheorem{assumption}{Assumption}

\def\5n{\negthinspace \negthinspace \negthinspace \negthinspace \negthinspace }
\def\4n{\negthinspace \negthinspace \negthinspace \negthinspace }
\def\3n{\negthinspace \negthinspace \negthinspace }
\def\2n{\negthinspace \negthinspace }

\def\0{\mathbf{0}}
\def\1{\mathbf{1}}

\journal{Journal of Econometrics}

\begin{document}

\begin{frontmatter}

\title{Forecasting security's volatility using low-frequency historical data, high-frequency historical data and option-implied volatility}

\author[a]{Huiling Yuan}
\author[b,c]{Yong Zhou}
\author[d]{Zhiyuan Zhang}
\author[d]{Xiangyu Cui\corref{cor1}}

\cortext[cor1]{Corresponding author: Xiangyu Cui. Address: 777, Guoding Road, Yangpu, Shanghai, China. Tel: 86-21-65904311. Fax: 86-21-65901055. Email: cui.xiangyu@mail.shufe.edu.cn.}
\address[a] {School of Statistics and Information, Shanghai University of International Business and Economics.}
\address[b] {Institute of Statistics and Interdisciplinary Sciences and School of Statistics, Faculty of Economics and Management, East China Normal University.}
\address[c] {Academy of Mathematics and Systems Sciences, Chinese Academy of Sciences.}
\address[d] {School of Statistics and Management, Shanghai University of Finance and Economics.}

\begin{abstract}
Low-frequency historical data, high-frequency historical data and option data are three major sources, which can be used to forecast the underlying security's volatility. In this paper, we propose two econometric models, which integrate three information sources. In GARCH-It\^{o}-OI model, we assume that the option-implied volatility can influence the security's future volatility, and the option-implied volatility is treated as an observable  exogenous variable. In GARCH-It\^{o}-IV model, we assume that the option-implied volatility can not influence the security's volatility directly, and the relationship between the option-implied volatility and the security's volatility is constructed to extract useful information of the underlying security. After providing the quasi-maximum likelihood estimators for the parameters and establishing their asymptotic properties, we also conduct a series of simulation analysis and empirical analysis to compare the proposed models with other popular models in the literature. We find that when the sampling interval of the high-frequency data is 5 minutes, the GARCH-It\^{o}-OI model and GARCH-It\^{o}-IV model has better forecasting performance than other models.
\end{abstract}

\bigskip


\end{frontmatter}

\section{Introduction}
Forecasting the volatility of a financial security is a very important topic in modern financial practice. The natural information source of the volatility is the historical data of the security, which can be further divided into low-frequency historical data and  high-frequency historical data. The low-frequency historical data are referred to as observed historical price data on the security at daily or longer time horizons.  The autoregressive conditional heteroskedasticity (ARCH) model proposed in \cite{Engle:1982} and the generalized autoregressive conditional heteroskedasticity (GARCH) model proposed in \cite{Bollers:1986} are the most famous models for the analysis of low-frequency historical data.  The high-frequency historical data are referred to as the intra-day historical price data on the security, such as tick-by-tick data, 1-second data, 5-minute data and etc. The scholars often model the high-frequency historical data by continuous-time It\^{o} processes and develop realized volatility estimators. These estimators include two-time scale realized volatility (TSRV) (\citealp{Zh:2005}), multi-scale realized volatility (MSRV) (\citealp{Zh:2006}), kernel realized volatility (KRV) (\citealp{B:2008}), pre-averaging realized volatility (PRV) (\citealp{J:2009}) and quasi-maximum likelihood estimator (QMLE) (\citealp{X:2010}), among others.

If there are options on the security in the market, the options' price data is another important information source of the security's volatility. The most famous and important volatility information extracted from the options' price data is the option-implied volatility (IV), which, when plugged into an option pricing model (e.g., the Black-Scholes model), returns a theoretical value equal to the current market price of an option, or is a weighted sum of the real-time, mid-quote prices of out-of-money call and put options (e.g., the CBOE Volatility Index, VIX).\footnote{\cite{CarrWu} showed that the VIX squared approximates the conditional risk-neutral expectation of the annualized return variance of S\&P 500 over the next 30 calendar days. Thus, we prefer to term such volatility index as option-implied volatility too. Since introduced in 1993, the VIX Index has been considered as the world premier barometer of investor sentiment and market volatility.} The option-implied volatility reflects the market's expectation of the security's future volatility, which has a good prediction power (\citealp{Chiras:1978}; \citealp{Beckers:1981}). Different from forecasting the volatility by using the historical data, which is backward looking, forecasting volatility by using the option-implied volatility is forward looking.

In the literature, there are some works that try to integrate two of the three major information sources, i.e., low-frequency historical data, high-frequency historical data and option data. On the one hand, there are several famous econometric models combining the low-frequency and high-frequency historical data, such as the realized GARCH model (\citealp{Hansen:2012}), the high-frequency-based volatility model (\citealp{Shephard:2010}), the multiplicative error model (\citealp{Engle:2006}), the Heterogenous Autoregressive model for Realized Volatility (HAR-RV) (\citealp{Corsi:2009}), the Mixed Data Sampling model (MIDAS) (\citealp{Ghysels:2006}), the GARCH-It\^{o} model (\citealp{KW:2016}) and the factor GARCH-It\^o model (\citealp{KimFan}), among others. The first three models integrate the daily realized volatility estimators into the low-frequency econometric models as exogenous variables. The fourth and fifth models construct linear models for the realized volatility estimators. The last two GARCH-It\^{o} models are a GARCH model for the low-frequency historical data at the integer time points and a continuous-time It\^{o} process model for the high-frequency historical data between the integer time points. Thus, compared with the other models, the GARCH-It\^{o} model provides more detailed structure for the high-frequency historical data and may make more efficient usage of the high-frequency historical data.

On the other hand, there are also several works combining the low-frequency historical data and the option-implied volatility.\footnote{There are also some works using other option information indexes instead of option-implied volatility. For example, \cite{Ni:2008} and \cite{Chang:2010} integrated the net demand for volatility of the non-market makers and the foreign institutional investors into linear volatility forecasting models. \cite{LW:2013} assumed the Black-Scholes model holds and treated the observed option price as a measurement of the theoretical one.}  \cite{Blair:2001} and  \cite{Ko:2005} integrated option-implied volatility into ARCH and GARCH models as exogenous variable. Differently, \cite{HaoZhang} and \cite{Ka:2014} derived the theoretical VIX value under the GARCH model and considered the observed VIX as a measurement of the theoretical one.

In this paper, we propose two econometric models, GARCH-It\^{o}-OI model and GARCH-It\^{o}-IV model, which are based on \cite{KW:2016}'s GARCH-It\^o model and integrate low-frequency historical data, high-frequency historical data and option data. In GARCH-It\^{o}-OI model, we assume that the investors would like to adjust their evaluation on the security's future volatility and their investment decision according to the observed option-implied volatility, which reflects the market's expectation on the security's future volatility. Then, the option-implied volatility is considered as an exogenous variable, which can influence the security's future volatility directly. In GARCH-It\^{o}-IV model, we assume that the options are redundant assets, which only contains useful information of the dynamics of the underlying security, but can not influence the security's volatility directly. Then, the relationship between the option-implied volatility and the security's volatility is constructed to extract useful information of the underlying security from the observed option-implied volatility. After proposing two models, we obtain the quasi-maximum likelihood estimators for the parameters and establish their asymptotic properties, respectively. We also conduct simulation studies for three different theoretical volatility models, Hesten model, Jump-diffusion model and  GARCH-It\^o-OI model,  and empirical studies for three different securities, S\&P 500 index future, APPLE stock and Sugar future. Through these studies, we compare the forecasting power of the proposed volatility models with other popular models in the literature, such as HAR-RV, HAR-RV-OI, Realized GARCH(RV), Realized GARCH-OI, Realized GARCH(IV), GARCH+OI, and GARCH-It\^{o}. We find that when the sampling interval of the high-frequency data is 5 minutes, the proposed GARCH-It\^{o}-OI model and GARCH-It\^{o}-IV model have stronger forecasting power.  However, when the sampling interval of the high-frequency data is 1 minute or 10 seconds, the HAR-RV model has stronger forecasting power. These findings suggest that for the models of GARCH-It\^o type, specifying the dynamic structure of the high-frequency data as an It\^o process with time changing volatility may not be accurate and helpful when the sampling interval is small. Thus, how to model an explicit mixed-frequency model when the the sampling interval of the high-frequency data is small is an interesting and changeling future research work.

This paper is organized as follows. Section 2 proposes the GARCH-It\^{o}-OI model and the GARCH-It\^{o}-IV model. Section 3 compares the GARCH-It\^{o}-OI model and GARCH-It\^{o}-IV model with other volatility models in the literature from the modelling aspect. Section 4 derives the quasi-maximum likelihood estimators for the parameters and develops asymptotic properties. Section 5 provides the simulation studies to compare the prediction performance of the proposed models with other models under different theoretical volatility assumptions. Section 6 provides the detailed empirical studies to illustrate the forecasting power of the proposed models for different underlying securities. Section 7 gives the conclusion and possible future works. All the proofs are given in the online Appendix.

\section{Unified models combining high-frequency historical data, low-frequency historical data and option-implied volatility}

\subsection{GARCH-It\^{o}-OI model}
The basic building block of our proposed models is \cite{KW:2016}'s GARCH-It\^{o} model, which embeds a standard GARCH(1,1) model into a continuous-time It\^{o} process and is a unified explicit model of low-frequency historical data and high-frequency historical data. More specifically, GARCH-It\^{o} model reads,
\begin{align*}
dX_{t}=&~\mu dt+\sigma_{t} dB_{t},\\
\sigma_{t}^{2}=&~\sigma_{[t]}^{2}+(t-[t])\left\{\omega+(\gamma-1)\sigma_{[t]}^{2}\right\}+\beta\left(\int_{[t]}^{t}\sigma_{s}dB_{s}\right)^{2},
\end{align*}
where $X_t$ is the log security price, $\mu$ is a drift, $[t]$ denotes the integer part of $t$, $B_{t}$ is a standard Brownian motion with respect to a filtration $\mathcal{F}_t$, $\sigma_t^2$ is a volatility process adapted to $\mathcal{F}_t$. \cite{KW:2016} further showed that under GARCH-It\^o model with $\mu=0$, the conditional variance of the log security price at integer time point $(n-1)$, $h_n$, follows a GARCH(1,1) structure,
\begin{align}
\label{eq:garch-ito-discrete}
h_{n}=w^g + \gamma h_{n-1}+\beta^g Z_{n-1}^2,
\end{align}
where $Z_{n-1}=X_{n-1}-X_{n-2}$ is the low-frequency log return, $\omega^{g}=\beta^{-1}(e^{\beta}-1)\omega$, $\beta^{g}=\beta^{-1}(\gamma-1)(e^{\beta}-1-\beta)+e^{\beta}-1$.

Now, we would like to integrate the option information (i.e., option-implied volatility) into the GARCH-It\^o model. We assume that the options are not redundant assets and the trading activities of the options may influence the prices of underlying security. Thus, the investors may adjust their evaluation on the security’s future volatility and their corresponding investment decision according to the observed option-implied volatility. To describe such circumstance, we add the observable option-implied volatility into the dynamic equation of instantaneous volatility as an exogenous variable. More specifically, we define the GARCH-It\^{o}-OI model as follows.

\begin{definition}\label{def_1}
We call a log security price $X_t$, $t\in[0,+\infty)$, to follow a GARCH-It\^{o}-OI model, if it satisfies
\begin{align*}
 dX_{t}=&\mu dt+\sigma_{t}dB_{t},\\
 \sigma_{t}^{2}=&\sigma_{[t]}^{2}+(t-[t])\left\{\omega+(\gamma-1)\sigma_{[t]}^{2}+\alpha O_{[t]}\right\}+\beta\left(\int_{[t]}^{t}\sigma_{s}d B_{s}\right)^{2}.
\end{align*}
where $\mu$ is a drift, $[t]$ denotes the integer part of $t$, $B_{t}$ is a standard Brownian motion with respect to a filtration $\mathcal{F}_t$, $\sigma_t^2$ is a volatility process adapted to $\mathcal{F}_t$ and $O_{[t]}$ is the $\mathcal{F}_{[t]}$-adapted exogenous option-implied variance at integer time $[t]$.
\end{definition}

We further denote the model parameters as $\theta=(\omega,\beta,\gamma,\alpha)^T$, where the superscript $T$ is used to denote the transpose of a matrix or a vector.  As $\sigma_{[t]}^2$ represents the historical information  and $O_{[t]}$ represents the option information, which have backward looking property and forward looking property, respectively. We may expect that $\sigma_{[t]}^2$ ($O_{[t]}$) has less (larger) influence on $\sigma_t^2$ as long as $t$ increases, which implies that $0<\gamma<1$ ($\alpha>0$).  Furthermore, according to Proposition 1 of online appendix, under GARCH-It\^o-OI model with $\mu=0$, the conditional variance of the log security price at integer time point $(n-1)$, $h_n$, also follows a GARCH(1,1) structure,
\begin{align}
\label{eq:garch-ito-oi-discrete}
h_{n}=w^g + \gamma h_{n-1}+\beta^g Z_{n-1}^2 + \eta^g O_{n-1} +\xi^g O_{n-2},
\end{align}
where $Z_{n-1}=X_{n-1}-X_{n-2}$ is the low-frequency log return,
$\omega^{g}=\beta^{-1}(e^{\beta}-1)\omega$, $\beta^{g}=\beta^{-1}(\gamma-1)(e^{\beta}-1-\beta)+e^{\beta}-1$, $\eta^{g}=\beta^{-2}(e^{\beta}-1-\beta)\alpha$, $\xi^{g}=\left[\beta^{-1}(e^{\beta}-1)-\beta^{-2}(e^{\beta}-1-\beta)\right]\alpha$.

\subsection{GARCH-It\^{o}-IV model}
In this subsection, we assume that the options are redundant assets and the trading activities of the options may not influence the prices of underlying security directly. To model such circumstance, we derive an additional relationship equation, which describes how the option-implied volatility depends on the security's volatility.

Now, we focus on the VIX-type option-implied volatility, $IV_t^{Model}$, which denotes the annualized option-implied volatility of a given underlying security over the time interval $[t,T]$.\footnote{Actually, $IV_t^{Model}$ is defined by the following formula,
\begin{align*}
(IV_t^{Model})^2=\frac{2}{T-t} \sum_{i} \frac{\Delta K_i}{K_i^2} e^{r_t(T-t)}O_{t}(K_i,T)-\frac{1}{T-t}\left[\frac{F_t}{K_0}-1\right],
\end{align*}
where $T$ is the expiry date for the options, $\Delta K_i = (K_{i+1} - K_i )/2$, $K_i$ is the strike price of the $i$th out-of-the-money option, $r_t$ denotes the time-$t$ risk-free rate with maturity $T$, $O_t(K_i,T)$ denotes the time-$t$ mid-quote price of the option,  $F_t$ is the time-$t$ forward price derived from coterminal option prices and $K_0$ is the first strike below the forward price $F_t$.} As shown in \cite{CarrWu}, the VIX-type option-implied volatility squared approximates the risk-neutral expectation of the annualized return variance from time $t$ to $T$,
\begin{align*}
(IV_t^{Model})^2 \approx E^Q [ h_{t,T} |\mathcal{F}_t],
\end{align*}
where $E^Q[\cdot]$ denotes the risk-neutral expectation, $\mathcal{F}_{t}$ is the information set at  time $t$. \cite{HaoZhang} and \cite{Ka:2014} further showed when the security's return follows a GARCH(1,1) model under the locally risk-neutral probability measure $Q$ proposed by \cite{Duan}, we have
\begin{align*}
(IV_t^{Model})^2 \approx  E^Q [ h_{t,T} |\mathcal{F}_t] = a h_{t+1} + b,
\end{align*}
where $h_{t+1}$ is the conditional variance, parameters $a$ and $b$ depend on the parameters of GARCH(1,1) model. There is a difference $e_t$ between the observed option-implied variance $IV_t^2$ and the model derived implied variance $(IV_t^{Model})^2$  as follows,
\begin{align}\label{eq:error}
IV_t^2= a h_{t+1} + b+e_t.
\end{align}
The difference is often assumed to be a white noise process $\{e_t\}\sim i.i.d. N(0,\sigma_e^2)$ (\citealp{HaoZhang}), or an autoregressive process, $e_{t+1}=\rho e_t + u_t$, where $\{u_t\}\sim i.i.d. N(0,\sigma_u^2)$ (\citealp{Ka:2014}). When $\{e_t\}$ is an autoregressive process, equation $(\ref{eq:error})$ becomes
\begin{align}\label{eq:dynamic-vix}
IV_t^2= \rho IV_{t-1}^2 + a h_{t+1} - \rho a h_t +b (1-\rho) +u_t,
\end{align}
where $\{u_t\}\sim i.i.d. N(0,\sigma_u^2)$. Equation (\ref{eq:dynamic-vix}) is the required relationship equation, which describes how the option-implied volatility depends on the security volatility.

 As revealed in equation (\ref{eq:garch-ito-discrete}), the low-frequency conditional variances of the GARCH-It\^{o} model obey a simple GARCH(1,1) structure.\footnote{This structure is still valid after changing the objective probability measure into the locally risk-neutral probability measure $Q$.} Thus, equation (\ref{eq:dynamic-vix}) also holds for the GARCH-It\^{o} model. We add equation (\ref{eq:dynamic-vix}) into the GARCH-It\^o model and obtain the GARCH-It\^{o}-IV model as follows.

\begin{definition}\label{def_2}
We call a log security price $X_t$, $t\in[0,+\infty)$, to follow a GARCH-It\^{o}-IV model, if it satisfies
\begin{equation*}
\begin{array}{rl}
 dX_{t}=\3n&\mu dt+\sigma_{t}dB_{t},\\[2mm]
 \sigma_{t}^{2}=\3n&\sigma_{[t]}^{2}+(t-[t])\left\{\omega+(\gamma-1)\sigma_{[t]}^{2}\right\}+\beta\left(\displaystyle\int_{[t]}^{t}\sigma_{s}d B_{s}\right)^{2},\\[3mm]
IV_{[t]}^2=\3n&\rho IV_{[t]-1}^2+a h_{[t]+1}-\rho a h_{[t]} + b(1-\rho) + u_{[t]},
\end{array}
\end{equation*}
where $\mu$ is a drift, $B_{t}$ is a standard Brownian motion with respect to a filtration $\mathcal{F}_t$, $\sigma_t^2$ is the instantaneous volatility process adapted to $\mathcal{F}_t$, $[t]$ denotes the integer part of $t$, $IV_{[t]}$ is the observed option-implied volatility at integer time $[t]$, $h_{[t]+1}$ is the conditional variance of daily return $X_{[t]+1}-X_{[t]}$, $u_{[t]}$ is a sequence of i.i.d. normal random variables with mean 0 and constant variance $\sigma_{u}^{2}$.
\end{definition}

We further denote the model parameters as $\theta=(\omega,\beta,\gamma)^T$, $\varphi=(\omega,\beta,\gamma,\rho, a, b)^T$ and $\phi=(\varphi^T,\sigma_u^2)^T$. In GARCH-It\^o-IV model, the option-implied volatility has an explicit dynamics, which is influenced by the latent conditional variance. Thus, GARCH-It\^{o}-IV model is a ``complete'' specification of the joint dynamics of return, latent instantaneous volatilities and option-implied volatility. Under GARCH-It\^o-IV model with $\mu=0$, the conditional variance of the log security price at integer time point $(n-1)$, $h_n$,  follows a GARCH(1,1) structure, and there is a stable relationship between the option-implied variance and the conditional variance,
\begin{align}
\label{eq:garch-ito-iv-1}
h_{n}&=w^g + \gamma h_{n-1}+\beta^g Z_{n-1}^2,\\
\label{eq:garch-ito-iv-2}
IV_n^2&=\rho IV_{n-1}^2 + a h_{n+1}-\rho a h_{n}+b(1-\rho) + u_n,
\end{align}
where $Z_{n-1}=X_{n-1}-X_{n-2}$ is the low-frequency log return,
$\omega^{g}=\beta^{-1}(e^{\beta}-1)\omega$, $\beta^{g}=\beta^{-1}(\gamma-1)(e^{\beta}-1-\beta)+e^{\beta}-1$.

\section{Comparison with other models}\label{section:3}
In this section, we would like to compare GARCH-It\^{o}-OI model, GARCH-It\^{o}-IV model with some of the other popular and quite related models in the literature from a modeling point of view. Although we have changed the notations of the models accordingly, there is still a slight abuse of notations in this comparison and we assume that no confusion will be caused in this section.

The first class of econometric models makes use of low-frequency historical data and option-implied volatility, which are low-frequency models and do not contain the continuous-time instantaneous variance process.

The GARCH+OI model in \cite{Ko:2005} assumes that the conditional variance of the security follows,
\begin{align*}
h_n=\omega+\gamma h_{n-1}+\beta Z_{n-1}^{2}+ \eta OI_{n-1}^2,
\end{align*}
which considers the option-implied volatility as an exogenous variable.

If we choose the option-implied volatility as the realized measure of volatility in the Realized GARCH model of \cite{Hansen:2012}, we obtain the Realized GARCH(IV) model, which assumes that the conditional variance of the security follows,
\begin{align*}
&h_n=\omega+\gamma h_{n-1}+\eta IV_{n-1}^2,\\
&IV_{n}^2=a h_{n+1}+ b +u_n,
\end{align*}
where $u_n \sim i.i.d.(0,\sigma_u^2)$.

In GARCH+OI model, the option-implied volatility has a power to shape the conditional variance, which implies that the options are not redundant. In Realized GARCH(IV) model, the option-implied volatility is a measure of the unobservable term $ah_{n+1} +b$, and can influence the dynamics of the conditional variance, which also implies that the options are not redundant. Thus, they are quite related to GARCH-It\^o-OI model.



The second class of econometric models makes use of low-frequency historical data and high-frequency historical data. In general, they involve the use of the realized volatility ($RV$) over a time interval, which is computed based on high-frequency historical data over that interval.

If we choose the realized volatility as the realized measure of volatility in the Realized GARCH model of \cite{Hansen:2012}, we obtain the Realized GARCH(RV) model, which assumes that the conditional variance of the security follows,
\begin{align*}
&h_n=\omega+\gamma h_{n-1}+\eta RV_{n-1}^2,\\
&RV_{n}^2=a h_{n+1}+ b +u_n,
\end{align*}
where $u_n \sim i.i.d.(0,\sigma_u^2)$.

The MIDAS method proposed in \cite{Ghysels:2006} is to find the best predictor of the future realized volatility by regressions among such possible measures of past fluctuations in returns as daily squared returns, absolutely daily returns, daily range, daily realized volatility, daily realized power variation and etc. Taking the daily realized volatility prediction as an example, we have
\begin{align*}
&RV_{n+1}=\mu+\rho \sum_{k=0}^{k_{max}} b(k,\theta)RV_{n-k}+\varepsilon_{n+1},
\end{align*}
where $RV_{n+1}$ is the realized volatility from time $n$ to time $n+1$, the lag coefficients $b(k,\theta)$ are parameterized as a function of a low-dimensional vector $\theta$. \cite{Ghysels:2006} suggested constructing $b(k,\theta)$ by Beta functions.

The HAR-RV model in \cite{Corsi:2009} is in a sense a MIDAS regression. It is derived from an economic Heterogeneous Market Hypothesis perspective, that is, different types of market participants have trading horizons at different frequencies, inducing different types of volatility components. The regression model for daily realized volatilities reads
\begin{align*}
RV^{(d)}_{n+1}=c+\beta^{(d)}RV_n^{(d)}+\beta^{(w)}RV_n^{(w)}+\beta^{(m)}RV_n^{(m)}+\varepsilon_{n+1}
\end{align*}
with $RV_n^{(d)}$, $RV_n^{(w)}$ and $RV_n^{(m)}$ are the realized volatilities over a day, a week and a month up to time $n$.

The GARCH-It\^{o} model in \cite{KW:2016} is a basic building block of our GARCH-It\^{o}-OI model and GARCH-It\^{o}-IV model. The corresponding low-frequency model of the conditional variance is
\begin{align*}
&h_n=\omega^{g}+\gamma h_{n-1}+\beta^{g} Z_{n-1}^{2}.
\end{align*}
To estimate the parameters, \cite{KW:2016} proposed the following quasi-likelihood function,
\begin{align*}
\hat{L}_{n,m}^{GH}(\theta) =-\frac{1}{2n} \sum_{i=1}^{n}\left(\log(h_i(\theta))+\frac{RV_i}{h_i(\theta)}\right),
\end{align*}
where $RV_i$ is the realized volatility based on high-frequency historical data during day $i$. The proposed quasi-likelihood function takes the same form as the likelihood function of daily log return, which follows the low-frequency GARCH model.

We can see that, under the Realized GARCH(RV), MIDAS, HAR-RV and GARCH-It\^{o} frameworks, the dynamics of conditional variance follows an AR model with order depending on the number of trading days involved.
More importantly, under the MIDAS and HAR-RV frameworks, the high-frequency  (continuous-time) and low-frequency (discrete-time) dynamic properties of the security prices are treated independently and only a low-frequency regression model is constructed. In GARCH-It\^{o}-OI model, the option-implied information is seen as an exogenous variable, and in GARCH-It\^{o}-IV model, the low-frequency dynamics of the option-implied volatility is further incorporated.
Moreover, the GARCH-It\^{o}-OI model and GARCH-It\^{o}-IV model could employ high-frequency historical data, low-frequency historical data and option-implied volatility in a more systematic way under a ``complete'' model.

\section{Parameter estimation for GARCH-It\^{o}-OI model and GARCH-It\^{o}-IV model}\label{section-4}
\subsection {Quasi-maximum likelihood estimators for GARCH-It\^{o}-OI model}\label{subsection-4.1}
Suppose that the underlying log price process $X_{t}$ follows the GARCH-It\^{o}-OI model in Definition \ref{def_1}. The low-frequency historical data are observed true log prices at integer times, namely $X_{i}, i= 0, 1, 2, \cdots, n$. The option-implied variances are computed at integer times based on the true prices of options, denoted as $O_{i}, i= 0, 1, 2, \cdots, n$. The high-frequency historical data are
observed log prices at time points between integer times, that is, $t_{i,j}$, $j=0,1,\cdots,m_i+1$, denote the high-frequency time points
during the $i$-th period satisfying $i-1=t_{i,0}<t_{i,1}<\ldots<t_{i,m_{i}}<t_{i,m_{i}+1}=t_{i+1,0}=i$. Different from the low-frequency historical data and the option-implied variances, the observed high-frequency log prices are contaminated by the micro-structure noise, and so the true high-frequency log prices are not observable. In light of this, we assume that observed high-frequency log prices $Y_{t_{i,j}}$ obey the simple additive noise model,
\begin{align*}
Y_{t_{i,j}}=X_{t_{i,j}}+\epsilon_{t_{i,j}},
\end{align*}
where $\epsilon_{t_{i,j}}$ is micro-structure noise independent of the process of $X_{t_{i,j}}$,
and for each $i$, $\epsilon_{t_{i,j}},j=1,\cdots,m_{i}$, are independent and identically distributed (i.i.d.) with mean zero and variance $\sigma_{\epsilon}^{2}$.

According to Proposition 1 of online Appendix, the conditional variance of the log security price at integer time point $(n-1)$, $h_n$, follows a GARCH(1,1) structure and is denoted as a function of model parameters, $g_n(\theta)$. Then, the log likelihood function for the low-frequency GARCH structure is given as follows,
\begin{align*}
l(\theta)
= -\frac{n}{2}\log(2\pi)-\frac{1}{2}\sum_{i=1}^{n}\log\left(g_{i}(\theta)\right)-\sum_{i=1}^{n}\frac{Z_{i}^{2}}{2(g_{i}(\theta))}.
\end{align*}

Similar to \cite{KW:2016}, we propose the quasi-likelihood function $\hat{L}_{n,m}^{GHO}$ for GARCH-It\^{o}-OI model as follows,
\begin{align*}
\hat{L}_{n,m}^{GHO}(\theta)=-\frac{1}{2n}\sum_{i=1}^{n}\log(g_{i}(\theta))-\frac{1}{2n}\sum_{i=1}^{n}\frac{RV_{i}}{g_{i}(\theta)},
\end{align*}
where the realized volatility $RV_{i}$ is estimated using $m_{i}$ high-frequency historical data during the $i$-th period and is treated as an ``observation''.\footnote{The estimation method of $RV_i$ is chosen among the multi-scale realized volatility estimator, preaveraging volatility realized estimator and kernel realized volatility estimator. The integrated volatility over the $i$-th period, $\int_{i-1}^{i}\sigma_t^2 dt$, equals to the sum of $g_i(\theta_0)$ and a martingale difference $D_i$, where $\theta_0 = (\omega_0, \beta_0, \gamma_0, \alpha_0)$ are the true values of the parameters. As the effects of martingale differences are asymptotically negligible, the martingale differences are dropped in the quasi-likelihood function $\hat{L}_{n,m}^{GHO}$. Please note that we need the initial values of $\sigma_{0}^{2}$ and $O_{0}$ to evaluate $\hat{L}_{n,m}^{GHO}(\theta)$.} We maximize the quasi-likelihood function $\hat{L}_{n,m}^{GHO}(\theta)$ over parameters'  space $\Theta$ and denote the maximizer as $\hat{\theta}^{GHO}$, that is,
 \begin{align*}
\hat{\theta}^{GHO}=\arg\max \limits_{\theta\in\Theta}\hat{L}_{n,m}^{GHO}(\theta).
\end{align*}
$\hat{\theta}^{GHO}= (\omega^{GHO}, \beta^{GHO}, \gamma^{GHO}, \alpha^{GHO})^T$ are the quasi-maximum likelihood estimators of $\theta_0 = (\omega_0, \beta_0, \gamma_0, \alpha_0)^T$.

\subsection {Quasi-maximum likelihood estimators for GARCH-It\^{o}-IV model}\label{subsection-4.2}
As shown in equation (\ref{eq:garch-ito-iv-1}) and (\ref{eq:garch-ito-iv-2}),  the conditional variance of the log security price at integer time point $(n-1)$, $h_n$,  follows a GARCH(1,1) structure, and there is a stable relationship between the option-implied variance and the conditional variance. We can write the joint quasi-likelihood function for the low-frequency daily log return and the option-implied volatility as follows,
\begin{align*}
&L\left(Z_{n},IV_{n-1},Z_{n-1},\cdots,IV_{1},Z_{1},IV_{0}\right)\\
=& f\left(Z_{n}|IV_{n-1},Z_{n-1},\cdots,IV_{1},Z_{1},IV_0\right) \cdot  f\left(IV_{n-1}|Z_{n-1},\cdots,IV_{1},Z_{1},IV_0\right)\cdots\\
\cdot&  f\left(Z_{2}|IV_{1},Z_{1},IV_0\right)\cdot f\left(IV_{1}|Z_{1},IV_0\right)\cdot f\left(Z_{1}|IV_0\right)\\
=&\prod_{i=1}^{n}\frac{1}{\sqrt{2\pi h_{i}}}\exp\left\{-\frac{Z_{i}^{2}}{2h_{i}}\right\} \cdot \prod_{j=1}^{n-1}\frac{1}{\sqrt{2\pi\sigma_{u}^{2}}}\exp\left\{-\frac{\left(IV_{j}^2-f_j(\varphi)\right)^{2}}{2\sigma_{u}^{2}} \right\},
\end{align*}
where
\begin{align}\label{eq:f}
f_j(\varphi):=\rho IV_{j-1}^2+a h_{j+1}(\theta)-\rho a h_{j}(\theta)+b(1-\rho).
\end{align}
 Thus, the log joint quasi-likelihood function is,
\begin{align*}
&l\left(Z_{n},IV_{n-1},Z_{n-1},\cdots,IV_{1},Z_{1},IV_{0}\right)\\
=&-\frac{2n-1}{2}\log2\pi-\frac{1}{2}\sum_{i=1}^{n}\left(\log h_{i}+\frac{Z_{i}^{2}}{h_{i}}\right)-\frac{1}{2}\sum_{j=1}^{n-1}\left(\log\sigma_{u}^{2}+\frac{\left(IV_{j}^2-f_j(\varphi) \right)^{2}}{\sigma_{u}^{2}}
\right).
\end{align*}

Similar to \cite{KW:2016}, we propose the quasi-likelihood function $\widetilde{L}_{n,m}^{GHO}$ for GARCH-It\^{o}-IV model as follows,
\begin{align*}
\widetilde{L}_{n,m}^{GHO}(\phi)=&-\frac{1}{2n}\sum_{i=1}^{n}\left(\log g_{i}(\theta)+\frac{RV_{i}}{g_{i}(\theta)}\right)-\frac{1}{2n}\sum_{j=1}^{n-1}\left(\log\sigma_u^2+\frac{(IV_{j}^{2}-f_j(\varphi))^{2}}{\sigma_u^2}\right),
\end{align*}
where the conditional variance $h_n$ is denoted as a function of model parameters $g_n(\theta)$, which only depends on the first three parameters of GARCH-It\^{o}-IV model.   We can see that the current joint quasi-likelihood function contains the conditional quasi-likelihood for the low-frequency GARCH structure and the conditional likelihood for the option-implied volatility error terms, and treats the realized volatilities as low-frequency ``observations''. The first part captures the high-frequency historical information, while the second part carries the added information from the option-implied volatility. Please note that we need the initial value $\sigma_{0}^{2}$ to evaluate $\widetilde{L}_{n,m}^{GHO}(\phi)$ and choose $\sigma_0^2=Z_1^2$. We maximize the quasi-likelihood function $\widetilde{L}_{n,m}^{GHO}(\phi)$ over parameters'  space $\Phi$ and denote the maximizer as $\tilde\phi^{GHO}$, that is,
 \begin{align*}
\tilde\phi^{GHO}=\arg\max \limits_{\phi\in\Phi}\widetilde{L}_{n,m}^{GHO}(\phi).
\end{align*}
$\tilde\phi^{GHO}= (\omega^{GHO}, \beta^{GHO}, \gamma^{GHO}, \rho^{GHO}, a^{GHO}, b^{GHO},(\sigma_u^2)^{GHO})^T$ are the quasi-maximum likelihood estimators of real parameters $\phi_0 = (\omega_0, \beta_0, \gamma_0, \rho_{0}, a_{0}, b_{0},(\sigma_{u}^2)_0)^T$.

\subsection {Asymptotic theory of estimators}
In this subsection, we try to establish consistency and asymptotic distribution for the proposed estimators $\hat{\theta}^{GHO}= (\omega^{GHO}, \beta^{GHO}, \gamma^{GHO}, \alpha^{GHO})^T$ for GARCH-It\^{o}-OI model and $\tilde\phi^{GHO}= \left(\omega^{GHO}, \beta^{GHO}, \gamma^{GHO}, \rho^{GHO}, a^{GHO}, b^{GHO},(\sigma_u^2)^{GHO}\right)^T$ for GARCH-It\^{o}-IV model.

First, we fix some notations. For a matrix $A = (A_{i,j})_{i,j=1,\dots,k}$, and a vector $a = (a_1, \dots , a_k)$, define $||A||_{\max} = \max_{i,j} |A_{i,j}|$ and $||a||_{\max} = \max_{i} |a_i |$. Given a random variable $X$ and $p \geq 1$, let $||X||_{L_p}= \{E [ |X|^{p}]\}^{1/p}$. Let $C$ be positive generic constants whose values are free of $\theta$, $\phi$, $n$ and $m_i$, and may change from appearance to appearance. Then, we give the following assumptions, under which the asymptotic theory is established.
\begin{assumption}\label{assumption}
(a1) In  GARCH-It\^{o}-OI model, let
\begin{align*}
\Theta=\{\theta=(\omega,\beta,\gamma,\alpha)^T~|~\omega_{l}<\omega<\omega_{u}, \beta_{l}<\beta<\beta_{u}, \gamma_{l}<\gamma<\gamma_{u},
\alpha_{l}<\alpha<\alpha_{u},\gamma+\beta^{g}<1\},
\end{align*}
where $\omega_{l},\omega_{u}, \beta_{l},\beta_{u}, \gamma_{l},\gamma_{u}, \alpha_{l},\alpha_{u}$ are known positive constants, and $\beta^{g}=\beta^{-1}(\gamma-1)(e^{\beta}-1-\beta)+e^{\beta}-1$.

(a2) In GARCH-It\^{o}-IV model, let
\begin{align*}
\Phi=&\{\phi=(\omega,\beta,\gamma, \rho, a, b, \sigma_{u}^2 )^T~|~\omega_{l}<\omega<\omega_{u}, ~\beta_{l}<\beta<\beta_{u},~ \gamma_{l}<\gamma<\gamma_{u},
~|\rho|<1\\
~&~a_{l}<a<a_{u},~ b_{l}<b<b_{u},~ (\sigma_{u}^2)_l<\sigma_{u}^2<(\sigma_{u}^2)_u, ~\gamma+\beta^g<1\},
\end{align*}
where $\omega_{l}$, $\omega_{u}$, $\beta_{l}$, $\beta_{u}$, $\gamma_{l}$, $\gamma_{u}$,$a_{l}$, $a_{u}$,  $b_{l}$, $b_{u}$, $(\sigma_{u}^2)_l$, $(\sigma_{u}^2)_u$ are known constants, $\beta^{g}=\beta^{-1}(\gamma-1)(e^{\beta}-1-\beta)+e^{\beta}-1$.

(b1) In GARCH-It\^{o}-OI model, the option-implied variance $\{O_{n}\geq 0: i\in\mathbb{N}\}$ is uniformly bounded.

(b2) In GARCH-It\^{o}-IV model, for any given $i\in \mathbb{N}$, $D_{i}$ and $u_{i}$ are independent.

(c1) $\frac{E\left[Z_{i}^{4}|\mathcal{F}_{i-1}\right]}{g_{i}^{2}(\theta_{0})}\leq C$ a.s. for any $i\in \mathbb{N}$.

(c2) There exists a positive constant $\delta$ such that $E\left[\left(\frac{Z_{i}^{2}}{g_{i}(\theta_{0})}\right)^{2+\delta}\right]\leq C$ for $i\in \mathbb{N}$.

(d) $\left\{|D_{i}| ~|~ i \in \mathbb{N}\right\}$ is uniformly integrable.

(e1) For GARCH-It\^{o}-OI model, $(D_{i},Z_{i}^{2})$ is a stationary ergodic process.

(e2) For GARCH-It\^{o}-IV model, $(D_{i},Z_{i}^{2},u_i)$ is a stationary ergodic process.

(f) Let $m=\sum_{i=1}^{n}m_{i}/n$. We have $C_{1}m\leq m_{i} \leq C_{2}m$, $\sup\limits_{1\leq j\leq m_{i}}|t_{i,j}-t_{i,j-1}|=O(m^{-1})$ and $n^{2}m^{-1}\rightarrow 0$ as
$m, n\rightarrow \infty$.

(g) $\sup\limits_{i\in \mathbb{N}}\left\|RV_{i}-\int_{i-1}^{i}\sigma_{t}^{2}dt\right\|_{L_{1+\delta}}\leq C \cdot m^{-1/4}$ for some $\delta >0 $.

(h) For any $i\in \mathbb{N}$, $E\left[RV_{i}|\mathcal{F}_{i-1}\right]\leq C\cdot E[\int_{i-1}^{i}\sigma_{t}^{2}dt|\mathcal{F}_{i-1}]+C$ a.s.
\end{assumption}

Comparing to the Assumption 1 in \cite{KW:2016}, we add additional Assumptions (b1) and (b2), Assumption (b1) is for the option-implied variance in GARCH-It\^o-OI model, and Assumption (b2) is for $D_i$ and $u_i$ in GARCH-It\^o-IV model. Among these assumptions, Assumption (a1)-(e2) are for the low-frequency part of the model and Assumption (f)-(h) are for the high-frequency part of the model. \cite{KW:2016} had explained that Assumption (c)-(h) are reasonable.  Assumption (b1) and (b2) are easily satisfied. Thus, these assumptions are all reasonable.

The following Theorem 1 and 2 establish the asymptotic theories  for $\hat \theta^{GHO}$ of GARCH-It\^o-OI model. The Theorem 3 and 4 establish the asymptotic theories  for $\tilde\phi^{GHO}$ of GARCH-It\^o-IV model.

\begin{theorem}\label{theorem-1}
(a) Under Assumption \ref{assumption} (a1), (b1), (d), (f)-(g), there is a unique maximizer of $L_{n}^{GHO}(\theta)$ and as $m,n \rightarrow\infty$,  $\hat{\theta}^{GHO}\rightarrow
\theta_{0}$ in probability, where
$$L_{n}^{GHO}(\theta)=-\frac{1}{2n}\sum_{i=1}^{n}\log(g_{i}(\theta))-\frac{1}{2n}\sum_{i=1}^{n}\frac{g_{i}(\theta_{0})}{g_{i}(\theta)}.$$

(b) Under Assumption \ref{assumption} (a1), (b1), (c)-(d), (f)-(h), we have
\begin{align*}
\left\|\hat{\theta}^{GHO}-\theta_{0}\right\|_{\max} = O_{p}\left(m^{-1/4}+n^{-1/2}\right).
\end{align*}
\end{theorem}

\begin{theorem}\label{theorem-2}
 Under Assumption \ref{assumption}, we have as $m, n\rightarrow \infty$,
\begin{align*}
 \sqrt{n}(\hat{\theta}^{GHO}-\theta_{0})\xrightarrow{d}N(0,B^{-1}A^{GHO}B^{-1}),
\end{align*}
where
\begin{align*}
A^{GHO}&=E\left[\frac{\partial g_{1}(\theta)}{\partial\theta}\frac{\partial
g_{1}(\theta)}{\partial\theta^{T}}\bigg|_{\theta=\theta_{0}}g_{1}^{-4}(\theta_{0})\int_{0}^{1}(e^{\beta_{0}(1-t)}-1)^{2}(X_{t}-X_{0})^{2}\sigma_{t}^{2}dt\right],\\
B&=\frac{1}{2}E\left[\frac{\partial g_{1}(\theta)}{\partial\theta}\frac{\partial g_{1}(\theta)}{\partial\theta^{T}}\bigg|_{\theta=\theta_{0}}g_{1}^{-2}(\theta_{0}) \right].
\end{align*}
\end{theorem}
Theorem 1 shows that $\hat{\theta}^{GHO}$ has the same convergence rate as the parameter estimators in GARCH-It\^{o} model of \cite{KW:2016}. In other words, the option-implied variance has no significant effect on the converge rate of the parameter estimators.

\begin{theorem}\label{theorem-3}
(a) Under Assumption \ref{assumption} (a2), (b2), (d), (f)-(g), there is a unique maximizer $\phi_0$ of $L_{n}^{GHO}(\phi)$ and as $m,n \rightarrow\infty$,  $\tilde\phi^{GHO}\rightarrow
\phi_{0}$ in probability, where
\begin{align*}
L_{n}^{GHO}(\phi)=&-\frac{1}{2n}\sum_{i=1}^{n}\left(\log g_{i}(\theta)+\frac{g_{i}(\theta_{0})}{g_{i}(\theta)}\right)-\frac{1}{2n}\sum_{j=1}^{n-1}\left\{\log\sigma_{u}^{2}+\frac{\left[f_j(\varphi)-f_j(\varphi_0)\right]^{2}+(\sigma_{u}^{2})_0}{\sigma_{u}^{2}}\right\}.
\end{align*}
(b) Under Assumption  (a2), (b2), (c)-(d), (f)-(h), we have
\begin{align*}
&\left\|\tilde\phi^{GHO}-\phi_{0}\right\|_{\max} = O_{p}\left(m^{-1/4}+n^{-1/2}\right),
\end{align*}
\end{theorem}

\begin{theorem}\label{theorem-4}
 Under Assumption \ref{assumption}, we have as $m, n\rightarrow \infty$,
\begin{align*}
 \sqrt{n}\left(\tilde{\phi}^{GHO}-\phi_{0}\right)\xrightarrow{d}N\left(0,\left(B^{GHO}\right)^{-1}A^{GHO}\left(B^{GHO}\right)^{-1}\right),
\end{align*}
where
\begin{align*}
A^{GHO}&=\begin{pmatrix}
A^{\varphi},&\0\\
\0^T&\frac{1}{2}((\sigma_{u}^2)_0)^{-2}
\end{pmatrix},\quad B^{GHO}=
\begin{pmatrix}
B^{\varphi}&\0\\
\0^T&\frac{1}{2}((\sigma_{u}^2)_0)^{-2}\\
\end{pmatrix},
\end{align*}
$A^{\varphi}$ and $B^{\varphi}$ are $6\times 6$ matrices as follows,
\begin{align*}
A^{\varphi}&=E\left[\frac{\partial g_{1}(\theta)}{\partial\varphi}\frac{\partial
g_{1}(\theta)}{\partial{\varphi}^{T}}\bigg|_{\varphi=\varphi_{0}} \frac{\int_{0}^{1}(e^{\beta_{0}(1-t)}-1)^{2}(X_{t}-X_{0})^{2}\sigma_{t}^{2}dt}{g_{1}^{4}(\theta_{0})} + \frac{\partial f_{1}(\varphi)}{\partial\varphi}\frac{\partial f_{1}(\varphi)}{\partial\varphi^{T}}\bigg|_{\varphi=\varphi_{0}}\frac{1}{(\sigma_{u}^2)_0}\right],\\
B^{\varphi}&=E\left[\frac{1}{2}\frac{\partial g_{1}(\theta)}{\partial\varphi}\frac{\partial g_{1}(\theta)}{\partial\varphi^{T}}\bigg|_{\varphi=\varphi_{0}}g_{1}^{-2}(\theta_{0}) + \frac{\partial f_{1}(\varphi)}{\partial\varphi}\frac{\partial f_{1}(\varphi)}{\partial\varphi^{T}}\bigg|_{\varphi=\varphi_{0}}\frac{1}{(\sigma_{u}^2)_0}\right],
\end{align*}
and $\0$ is 6-dimensional zero vector.
\end{theorem}
For the number of the observed option-implied volatilities is still $n$, $\tilde\phi^{GHO}$ has the same convergence rate as the estimators in GARCH-It\^{o} model of \cite{KW:2016}. We can also see that the second terms in $A^{\varphi}$ and $B^{\varphi}$ represent the influences of the dynamics of the option-implied volatility on the asymptotic variances of the estimations.

\section{Simulation study}
In this section, we study the prediction performance of the GARCH-It\^o-OI and GARCH-It\^o-IV model with different low-frequency sampling intervals: 1/4 day, 1/2 day and 1 day. And compare the GARCH-It\^o-OI and GARCH-It\^o-IV model with other models in the literature under different theoretical volatility models, Hesten model, Jump-diffusion model and GARCH-It\^o-OI model.

\subsection{Performance under different low-frequency sampling intervals.}
We consider a GARCH-It\^o-OI model with $\theta_{0}$=$(\omega_{0},\beta_{0},\gamma_{0},\alpha_{0})^T$=$(0.2,0.3,0.4,0.2,0.1)^T$, $X_0=10$, $\sigma_0^2=0.2$, $O_0\sim N(0,0.5)$ and $\epsilon_{t_{i,j}}\sim N(0,1e-6)$. The low-frequency sampling interval is 1/4 day and high-frequency sampling interval is 1/3120 day. We simulate data for 150 days, i.e., $150\times 4 \times 780$ log prices and $150\times 4$ option-implied variances. The $780^{\mbox{th}}$, $780\times 2^{\mbox{th}}$, $\cdots$, $780 \times 600^{\mbox{th}}$ log prices and the $1^{\mbox{st}}$, $2^{\mbox{nd}}$, $\cdots$, $600^{\mbox{th}}$ option-implied variances constitute the low-frequency data of 1/4 day low-frequency sampling interval. The $1560^{\mbox{th}}$, $1560\times 2^{\mbox{th}}$, $\cdots$, $1560 \times 300^{\mbox{th}}$ log prices and the $2^{\mbox{nd}}$, $2\times 2^{\mbox{th}}$, $\cdots$, $2\times 300^{\mbox{th}}$ option-implied variances constitute the low-frequency data of 1/2 day low-frequency sampling interval. And the $3120^{\mbox{th}}$, $3120\times 2^{\mbox{th}}$, $\cdots$, $3120 \times 150^{\mbox{th}}$ log prices and the $4^{\mbox{th}}$, $4\times 2^{\mbox{th}}$, $\cdots$, $4\times 150^{\mbox{th}}$ option-implied variances constitute the low-frequency data of 1 day low-frequency sampling interval. For GARCH-It\^o-IV model, we simulate sample paths similarly with the initial values of parameters being $\phi_{0}=(\omega_{0},\beta_{0},\gamma_{0},\rho_0, a_0, b_0, (\sigma_{u}^2)_0)^T=(0.2,0.3,0.4,0.2,0.1,0.001,0.04)^T$, $X_0=10$, $\sigma_0^2=0.2$, $IV_0^2=0.25$ and $\epsilon_{t_{i,j}}\sim N(0,1e-6)$.


We choose the last 50 days as out-of-sample period. For different low-frequency sampling intervals, we estimate the GARCH-It\^o-OI model and GARCH-It\^o-IV model based on the in-sample data and make predictions for the volatility of next day with a rolling horizon scheme. We define the following four criteria to evaluate the forecasting error, which are mean absolute error (MAE), mean square error (MSE), adjusted mean absolute percentage error (AMAPE) and logarithmic loss (LL),
\begin{align*}
&MAE=\frac{1}{N}\sum_{i=1}^{N}|RV_{i}-F_i|,\\
&MSE=\frac{1}{N}\sum_{i=1}^{N}(RV_{i}-F_i)^{2},\\
&AMAPE=\frac{1}{N}\sum_{i=1}^{N}\left|\frac{F_i-RV_{i}}{F_i+RV_{i}}\right|,\\
&LL=\frac{1}{N}\sum_{i=1}^{N}(\log(F_i)-\log(RV_{i}))^{2},
\end{align*}
where the realized volatility $RV_i$ is considered as the best estimation of the real integrated volatility in day $i$, $F_i$ is the volatility prediction in day $i$.

\begin{table}[!htb]
\caption{The prediction performance of the GARCH-It\^o-OI model and the GARCH-It\^o-IV model with different low-frequency sampling intervals.}
\label{table_numerical 1}
\begin{tabular}{ccccccccc}
\hline
\hline
& \multicolumn{4}{c}{GARCH-It\^{o}-OI}   &   \multicolumn{4}{c}{GARCH-It\^{o}-IV} \\
  \hline
 &MAE&MSE & AMAPE &   LL& MAE &MSE&  AMAPE &   LL \\
  \hline
1/4 day &0.3367& 0.2610&   0.0708 & 0.0335 &0.4162&0.3408 & 0.0888 & 0.0555 \\
1/2 day &0.6718&0.8536 & 0.1388& 0.1161 &0.7142& 0.8605& 0.1451 &  0.1281\\
1 day &0.8059&1.5382 & 0.1634 & 0.1751 &0.9748&2.1229&  0.1846&  0.2101\\
   \hline
   \hline
\end{tabular}
\end{table}

Table \ref{table_numerical 1} presents the forecasting errors of the GARCH-It\^o-OI model and the GARCH-It\^o-IV model with three different low-frequency sampling intervals: 1/4 day, 1/2 day, 1 day. We have two interesting findings. First, the GARCH-It\^o-OI model has better prediction performance than the GARCH-It\^o-IV model. Thus, given the same number of samples, the parameters of the GARCH-It\^o-OI model can be estimated more accurately. Second, the GARCH-It\^o-OI model and the GARCH-It\^o-IV model have better prediction performance, when the low-frequency sampling interval is 1/4 day. This may be partially because that i) the GARCH-It\^o-OI model and the GARCH-It\^o-IV model with low-frequency sampling interval being 1/4 day make use of more low-frequency data; and ii) the simulated data are respectively generated from the GARCH-It\^o-OI model and the GARCH-It\^o-IV model with low-frequency sampling interval being 1/4 day.

\subsection{Performance under different theoretical volatility models}
In this subsection, we simulate the sample data from three theoretical volatility models, Hesten model, Jump-diffusion model and GARCH-It\^o-OI model. Under each theoretical volatility model, we simulate three sample sets, whose high-frequency time intervals are 5 minutes, 1 minute and 10 seconds, respectively. We report the out-of-sample prediction performances of GARCH-It\^o model, GARCH-It\^o-OI model,  GARCH-It\^o-IV, Realized GARCH model and HAR-RV model.

\subsubsection{Heston stochastic volatility model}
We assume that the price of the security obeys the following Heston model,
\begin{align}
\begin{split}
dS(t)=&rS(t)dt+\sqrt{V(t)}S(t)dW_{1}(t),\\
dV(t)=&(a-bV(t))dt+\gamma\sqrt{V(t)}dW_{2}(t),
\end{split}
\end{align}
where $W_{1}(t), W_{2}(t)$ are Brownian motions with correlation coefficient $\rho$.
We further set $r=0$, $a=0.01$, $b=0.001$, $\gamma=0.075$, $\rho=-0.8$,  $S_0=50$, $V_0=0.05$. We consider three cases, where the high-frequency time intervals are 5 minutes, 1 minute and 10 seconds, respectively. And the volatility $V(t)$ generated from Heston model is considered as the option-implied volatility.  We run the simulation 1000 repetitions and there are 101 days in each repetition. The data in the first 100 days are used for estimating parameters, and the data in the 101st day is saved for out-of-sample testing. To compare the prediction performances of different models via more criteria, we introduce two more criteria as follows,
\begin{align*}
&HMAE=\frac{1}{N}\sum_{i=1}^{N}|1-RV_{i}^{-1}F_i|,\\
&HMSE=\frac{1}{N}\sum_{i=1}^{N}(1-RV_{i}^{-1}F_i)^{2}.
\end{align*}

\begin{table}[!htb]
{\small
\caption{The forecasting errors of different models under Heston model. }
\begin{tabular}{ccccccccc}
\hline
  \hline
&MAE& MSE & HMAE & HMSE & AMAPE &   LL&\\
  \hline
GARCH-It\^{o}(5 min)&1.9356e-04&6.8758e-08 &   0.5303 & 0.8365 &  0.2048&  0.3041\\
GARCH-It\^{o}-OI(5 min) &{\bf1.9188e-04}&{\bf6.6286e-08}&  {\bf 0.5284} & {\bf0.8059} & {\bf 0.2035}  & {\bf0.2969}  \\
GARCH-It\^{o}-IV(5 min)&{\bf1.9160e-04}&{\bf6.7935e-08} & {\bf  0.5202} &  {\bf 0.7841} & {\bf  0.2038}  & {\bf0.2989}  \\
HAR-RV(5 min)&1.9342e-04& 6.7061e-08 & 0.5399 & 0.8482 & 0.2040 &   0.3010\\
Realized GARCH(5 min) &2.2291e-04& 8.5601e-08& 0.7359 & 1.8357 & 0.2296 & 0.4018 \\
\hline
GARCH-It\^{o}(1 min)&1.4565e-04& 3.7544e-08&   0.2908 &  0.1589 &  0.1351&  0.1197\\
GARCH-It\^{o}-OI(1 min)& {\bf1.4411e-04}&{\bf3.5637e-08 }&  {\bf 0.2893} & {\bf0.1516} & {\bf 0.1345}  & {\bf0.1155}  \\
GARCH-It\^{o}-IV(1 min)& {\bf1.4767e-04}&{\bf4.0493e-08} & {\bf 0.2887} &  {\bf 0.1552} & {\bf  0.1368}  & {\bf0.1235}  \\
HAR-RV(1 min) &\text{\color{red}1.3861e-04}&\text{\color{red}3.3142e-08}& \text{\color{red}0.2779} & \text{\color{red}0.1456} & \text{\color{red}0.1271} & \text{\color{red}0.1042}\\
Realized GARCH(1 min)&1.7294e-04& 5.1838e-08& 0.3583 & 0.2618 & 0.1581 & 0.1898 \\

\hline
GARCH-It\^{o}(10 seconds) &1.223e-04&2.7613e-08&   0.2152 &  0.0813 &  0.1057&  0.0722\\
GARCH-It\^{o}-OI(10 seconds)&{\bf1.1215e-04}&{\bf2.2176e-08 }&  {\bf 0.1984} & {\bf0.0657} & {\bf 0.0969}  & {\bf0.0603}  \\
GARCH-It\^{o}-IV(10 seconds)&{\bf1.2831e-04}&{\bf2.9718e-08} & {\bf 0.2210} &  {\bf 0.0819} & {\bf  0.1109}  & {\bf0.0779}  \\
HAR-RV(10 seconds)&\text{\color{red}9.5073e-05}&\text{\color{red}1.5652e-08} & \text{\color{red}0.1727} & \text{\color{red}0.0527} &\text{\color{red} 0.0828} & \text{\color{red} 0.0448}\\
Realized GARCH(10 seconds)&1.4619e-04& 3.6053e-08& 0.2617 & 0.1144 & 0.1279 & 0.1139 \\

  \hline
   \hline
\end{tabular}
\label{table_error_heston_5 min, 1 min, 10 seconds}}
\end{table}

Table \ref{table_error_heston_5 min, 1 min, 10 seconds} displays the forecasting errors of different models. Among them, the models of  GARCH-It\^o type and HAR-RV model present strong forecasting power. We also find that when the high-frequency sample interval is 5 minutes, the GARCH-It\^o-OI model and GARCH-It\^o-IV model have the best prediction performances. That is because they include more data information. But, when the high-frequency sample interval is 1 minute or 10 seconds, the HAR-RV model has the best prediction performance. HAR-RV model is a simple linear model with different realized volatilities over different time periods. When the high-frequency sampling interval becomes small, the increasing high-frequency data makes the calculation of model more complex, and a simple model structure would have better performance. Thus, comparing to the simple structure of HAR-RV model, the dynamic structure of the high-frequency data in GARCH-It\^o type models may not be helpful in prediction for small high-frequency sampling intervals under the Heston model assumption. Furthermore, GARCH-It\^o-OI model shows better forecasting performance than GARCH-It\^o-IV model, due to a simpler structure of integrating option-implied information.

\subsubsection{Jump-diffusion model}
We assume that the price of the security obeys the following Jump-diffusion model,
\begin{align}
\begin{split}
dS(t)=&rS(t)dt+\sqrt{V(t)}S(t)dW_{1}(t)\\
dV(t)=&(a-bV(t))dt+\gamma\sqrt{V(t)}dW_{2}(t)+dJ_{t},
\end{split}
\end{align}
where $J_{t}=\sum_{i=1}^{N_{{t}}}Y_{i}$ is a compound Poisson process, $\{Y_{i}\}$ are i.i.d. $N(0,\sigma_{J}^{2})$ random variables, $\{N_t\}$ is a Poisson process with intensity $\lambda$. Besides the parameters in Heston model, we further set $\lambda=1$ and $\sigma_J=0.01$. We obtain the similar forecasting results as Subsection 5.2.1, which are represented in Table \ref{table_error__heston_jump_5 min, 1 min, 10 seconds}.

\begin{table}[!htb]
{\small
\caption{The forecasting errors of different models under Jump-diffusion model. }
\begin{tabular}{ccccccccc}
\hline
  \hline
&MAE& MSE & HMAE & HMSE & AMAPE &   LL&\\
  \hline
GARCH-It\^{o}(5 min) &2.2224e-04 &9.5106e-08&   0.5000 & 0.6848 &  0.2035&  0.2924\\
GARCH-It\^{o}-OI(5 min) & {\bf2.1693e-04}&{\bf9.0451e-08}&  {\bf 0.4991} & {\bf0.6967} & {\bf 0.1987}  & {\bf0.2846}  \\
GARCH-It\^{o}-IV(5 min) &{\bf2.1764e-04}&{\bf9.2604e-08}& {\bf  0.4775} &  {\bf 0.5798} & {\bf  0.2013}  & {\bf0.2819}  \\
HAR-RV(5 min) &2.2201e-04&8.8236e-08& 0.5204 & 0.7787 & 0.1999 &   0.2872\\
Realized GARCH(5 min)&2.4757e-04&1.0589e-07 & 0.6509 & 1.2632 & 0.2200 & 0.3641 \\
\hline
GARCH-It\^{o}(1 min) &1.8161e-04&5.9973e-08&   0.2837 &  0.1365 &  0.1412&  0.1275\\
GARCH-It\^{o}-OI(1 min)&{\bf1.7113e-04}& {\bf5.0757e-08}&  {\bf 0.2809} & {\bf0.1352} & {\bf 0.1347}  & {\bf0.1160}  \\
GARCH-It\^{o}-IV(1 min)&{\bf1.8170e-04}&{\bf 6.2671e-08}& {\bf 0.2758} &  {\bf 0.1252} & {\bf  0.1423}  & {\bf0.1315}  \\
HAR-RV(1 min)&\text{\color{red}1.6306e-04}&\text{\color{red}4.6206e-08} & \text{\color{red}0.2725} & \text{\color{red}0.1309} & \text{\color{red}0.1268} &   \text{\color{red}0.1025}\\
Realized GARCH(1 min) &1.9876e-04&6.9681e-08& 0.3345 & 0.2089 & 0.1573& 0.1695 \\

\hline
GARCH-It\^{o}(10 seconds)&1.2967e-04& 3.0154e-08&   0.1862 &  0.0556 &  0.0973&  0.0611\\
GARCH-It\^{o}-OI(10 seconds)&{\bf 1.1910e-04}&{\bf 2.4152e-08}&  {\bf 0.1803} & {\bf0.0543} & {\bf 0.0897}  & {\bf0.0519}  \\
GARCH-It\^{o}-IV(10 seconds) &{\bf 1.4228e-04}&{\bf 3.7157e-08}& {\bf 0.1957} &  {\bf 0.0587} & {\bf  0.1057}  & {\bf0.0714}  \\
HAR-RV(10 seconds) &\text{\color{red}1.0966e-04}&\text{\color{red}2.0200e-08}& \text{\color{red}0.1671}& \text{\color{red}0.0464} & \text{\color{red}0.0815} &\text{\color{red} 0.0420}\\
Realized GARCH(10 seconds) &1.5357e-04&4.0108e-08& 0.2306 & 0.0866 & 0.1185 & 0.1114 \\
   \hline
   \hline
\end{tabular}
\label{table_error__heston_jump_5 min, 1 min, 10 seconds}}
\end{table}

\subsubsection{GARCH-It\^{o}-OI Model}
We assume the log prices of the security follows GARCH-It\^{o}-OI Model. And we further set $\theta_{0}=(\omega_{0},\beta_{0},\gamma_{0},\alpha_{0})^T=(0.2,0.3,0.4,0.1)^T$. The exogenous option-implied variance $O_{n}$ is assumed to have a normal distribution $N(0,0.5)$.

The forecasting errors of different models is represented in Table \ref{table_error_GARCH_OI Normal Model_5 min, 1min, and 10 seconds}. We can see that for three different high-frequency sampling intervals,  the GARCH-It\^o-OI model and  GARCH-It\^o-IV model always have the best forecasting performances. The reason is that i) our proposed models make full use of the data information; ii) the sample data is generated by GARCH-It\^o-OI model. The dynamic structure of the high-frequency data assumed in GARCH-It\^o type models correctly describes the property of the sample data. Thus, the HAR-RV model can not beat the GARCH-It\^o type models any longer, when the high-frequency sampling interval is small.

\begin{table}[!htb]
{\small
\caption{The forecasting errors of different models under GARCH-It\^{o}-OI model.}
\begin{tabular}{ccccccccc}
\hline
  \hline
&MAE& MSE & HMAE & HMSE & AMAPE &   LL&\\
  \hline
GARCH-It\^{o}(5 min)&0.2573& 0.7199&   0.6376 &  1.2500 &   0.2219 & 0.3731 \\
GARCH-It\^{o}-OI (5 min)&{\bf0.2541}& {\bf0.6647}& {\bf 0.6326 }& {\bf1.3299} & {\bf 0.2203} & {\bf0.3688} \\
GARCH-It\^{o}-IV (5 min) &{\bf0.2609}&{\bf0.7251}& {\bf0.6540 }&   {\bf1.3041} & {\bf0.2240 }& {\bf 0.3802}\\
HAR-RV (5 min)&0.3133&0.9943& 0.8142 &  2.0206 &  0.2604 &    0.6858\\
Realized GARCH (5 min)&0.3724&0.9746& 1.1039 & 3.7967 & 0.2942& 0.8738 \\
\hline
GARCH-It\^{o} (1 min)&0.1954&0.14614&   0.3520 &  0.2297 &   0.1550 & 0.1622 \\
GARCH-It\^{o}-OI (1 min)&{\bf0.1941}&{\bf0.1437}& {\bf 0.3454 }& {\bf0.2135} & {\bf 0.1542} & {\bf0.1586} \\
GARCH-It\^{o}-IV (1 min)&{\bf0.1982}&{\bf 0.1532}& {\bf0.3616 }&   {\bf0.2446} & {\bf0.1565 }& {\bf 0.1654}\\
HAR-RV (1 min)&0.2505&0.1867& 0.4738 &  0.4201 &  0.1953&    0.4107\\
Realized GARCH (1 min)&0.2886&0.2124& 0.5695 & 0.6588 & 0.2201& 0.3935\\
\hline
GARCH-It\^{o} (10 seconds)&0.14850&0.0706&   0.2256 &  0.0879 &   0.1095 & 0.0869 \\
GARCH-It\^{o}-OI (10 seconds)&{\bf0.1456}&{\bf0.0687}& {\bf 0.2185 }& {\bf0.0819} & {\bf 0.1068} & {\bf0.0829} \\
GARCH-It\^{o}-IV (10 seconds)&{\bf0.1500}&{\bf0.0699}& {\bf0.2319 }&   {\bf0.0933} & {\bf0.1107 }& {\bf 0.0876}\\
HAR-RV (10 seconds)&0.2334&0.2153& 0.3484 &  0.2553 &  0.1563&    0.2608\\
Realized GARCH (10 seconds)&0.2837&0.2618& 0.4408 & 0.4074 & 0.1909& 0.3360\\
   \hline
   \hline
\end{tabular}
\label{table_error_GARCH_OI Normal Model_5 min, 1min, and 10 seconds}}
\end{table}

\section{Empirical study}
In this section, we use real trading data to compare the forecasting performances of our proposed models with other models in the literature. We consider three different securities: S\&P500 index future, APPLE stock, and Surge future.

\subsection{S$\&$P500 index future}\label{subsection 6.1}

The underlying security is S\&P500 index future. The high-frequency historical data is the 5-minute data from January 2, 2003 to December 31, 2012. In general, there are 78 prices in a trading day.  The low-frequency historical data is the daily close prices and the option-implied variances are the squared daily VIX index observations (multiplied by $1e-04$), ranging from January 2, 2003 to December 31, 2012. The number of analyzed high-frequency data is 194688, and the number of low-frequency data and VIX data is 2496. All high-frequency prices are transformed into log prices $\log (P_{t_{i,j}})$, $t_{i,j}=i-1+j/m$, $i=1,\ldots,n$, $j=1,\ldots,m$ with $n=2496$, $m=78$. The in-sample period starts from January 2, 2003 to December 31, 2007, which contains 97266 high-frequency prices and 1247 days. The out-of-sample period starts  from January 2, 2008 to December 31, 2012, which contains 97422 high-frequency prices and 1249 days. We can see that both the in-sample and out-of-sample periods contain a part of subprime mortgage crisis.

To illustrate the prediction power of the GARCH-It\^{o}-OI model and GARCH-It\^{o}-IV model more accurately, we also compute the forecasts of some valuable volatility models mentioned in Section \ref{section:3}, including GARCH+OI model,  Realized GARCH(IV) model, Realized GARCH(RV) model, HAR-RV model and GARCH-It\^{o} model. In addition, we introduce three new ones, Realized GARCH-OI model, HAR-RV-OI model and IV model. The Realized GARCH-OI model, HAR-RV-OI model are the extensions of Realized GARCH model and HAR-RV model, respectively, where the option-implied  variance is plugged in as an exogenous variable. IV model is a linear regression model based only on the option-implied variances as follows,
\begin{align*}
h_n=\omega+\beta_1 IV_{n-1}^2 +\beta_2 IV_{n-2}^2+u_n,
\end{align*}
where $\{u_n\}\sim i.i.d.N(0,\sigma_u^2)$. We also choose MAE, MSE, AMAPE and LL to evaluate the forecasting errors. From their definitions, we can see that AMAPE and LL measures give more weights on the small-scale volatilities. Table \ref{table_error 1 for SP} summarizes the forecasting errors of different models.\footnote{We need to point out that as there are negative volatility forecasts in IV model, LL of IV model is only computed for the positive volatility forecasts and thus is underestimated.}
\begin{table}[!htb]
\caption{The forecasting errors of different models for S\&P500 index future. }
\begin{tabular}{cccccc}
\hline
  \hline
  &MAE& MSE  & AMAPE &   LL\\
  \hline
GARCH-It\^{o} &6.1497e-05 &  2.2661e-08& 0.3083 & 0.7835 \\
GARCH-It\^{o}-OI &{\bf5.6657e-05} & {\bf1.9485e-08}& {\bf  0.2887} & {\bf 0.6759} \\
GARCH-It\^{o}-IV &{\bf6.2871e-05} & {\bf2.4445e-08}&   {\bf  0.3078} & {\bf 0.7686} \\
\hline
HAR-RV &6.2465e-05 &2.6640e-08 &   0.2916 &  0.7081\\
HAR-RV-OI &6.0882e-05 & 2.2319e-08&  0.2899 &  0.6909\\
Realized GARCH(RV) & 1.2499e-04& 4.4575e-08& 0.4628 &  1.4389 \\
Realized GARCH-OI &1.3879e-04 & 5.2464e-08& 0.4428 & 1.5349 \\

\hline
GARCH+OI &1.4564e-04 & 5.5123e-08 & 0.4989 & 1.9866 \\
Realized GARCH(IV) & 2.4495e-04& 1.4936e-07&  0.5854 & 3.3745 \\
\hline
IV  &6.330e-05 & 2.238e-08 & 0.2941 & 0.6763\\
   \hline
   \hline
\end{tabular}
\label{table_error 1 for SP}
\end{table}

Based on Table \ref{table_error 1 for SP}, we have the following interesting findings. First,  GARCH-It\^{o}-OI model has the stronger prediction power than other volatility models. Compared with GARCH-It\^o-IV model, the simple structure of GARCH-It\^{o}-OI model makes more efficient use of option-implied information. Second, although the HAR-RV-OI model has better prediction performance than HAR-RV model due to the option-implied information, it can not defeat the GARCH-It\^{o}-OI model. Third, Realized GARCH type models have the worst prediction performances. Fourth, IV model captures the stable linear relationship between the conditional variance and the option-implied variance and provides good prediction performance.

Furthermore, we also compare the prediction performances of different models during and after the subprime mortgage crisis. First, we select the data from January 2, 2003 to July 31, 2007 as in-sample data and the data during the subprime mortgage crisis (from August 1, 2007 to March 31, 2009) as out-of-sample data. Second, we select the data from January 2, 2003 to  December 31, 2010 as in-sample data and the data after the subprime mortgage crisis (from January 3, 2011 to December 31, 2012) as out-of-sample data.  The forecasting errors of different models are given in Table \ref{table_error_2 for SP}. We can see that GARCH-It\^o-OI model still has the best prediction performance. And most of models have better MAE and MSE, but worse AMAPE and LL after subprime mortgage crisis. After subprime mortgage crisis, the volatility of S\&P500 index future drops and there are more days with relative small daily volatilities (comparing to during subprime mortgage crisis). Most of models are weak to predict small daily volatilities.

\begin{table}[!htb]
\footnotesize
\caption{The forecasting errors during and after the subprime mortgage crisis for S\&P500 index future. }
\begin{tabular}{cccccccccc}
\hline
\hline
& \multicolumn{4}{c}{During subprime mortgage crisis} & \multicolumn{4}{c}{After subprime mortgage crisis}\\
  \hline
 &MAE & MSE  & AMAPE &   LL & MAE & MSE & AMAPE &   LL\\
  \hline
 GARCH-It\^{o} &1.126e-04&5.994e-08  & 0.286 & 0.614 &2.866e-05&2.987e-09& 0.298 & 0.776\\
GARCH-It\^{o}-OI& {\bf1.035e-04} & {\bf 5.070e-08} & {\bf0.272}& {\bf 0.560} & {\bf2.628e-05} & {\bf2.892e-09} & {\bf  0.278} & {\bf  0.636} \\
GARCH-It\^{o}-IV& {\bf1.183e-04}& {\bf6.557e-08}& {\bf 0.290} & {\bf 0.622} &{\bf2.881e-05}& {\bf3.187e-09} & {\bf  0.298} & {\bf 0.755} \\
\hline
HAR-RV &1.203e-04&7.148e-08& 0.278&  0.594&2.794e-05&3.729e-09&0.282 &  0.673\\
HAR-RV-OI &1.179e-04&5.904e-08& 0.277 &  0.572&2.694e-05& 3.033e-09& 0.278 &  0.643\\
Realized GARCH(RV)&1.608e-04&  6.438e-08 & 0.372 & 0.937 &1.087e-04& 4.645e-08 & 0.514& 1.721\\
Realized GARCH-OI& 2.849e-04&  2.275e-07& 0.466 & 1.682&5.464e-05&5.959e-09 & 0.442 & 1.582 \\
\hline
GARCH+OI & 2.071e-04 &  1.140e-07 & 0.404 & 1.250 & 1.036e-04 &  2.516e-08& 0.539 & 2.318\\
Realized GARCH(IV)  & 2.775e-04  & 2.405e-07 & 0.439 & 1.605   & 1.328e-04 & 2.070e-08& 0.652 & 3.891 \\
\hline
IV  & 1.249e-04& 5.914e-08 & 0.290& 0.628& 2.921e-05 & 3.254e-09 &  0.290 & 0.619\\
   \hline
   \hline
\end{tabular}
\label{table_error_2 for SP}
\end{table}

\subsection{APPLE stock }
We choose APPLE stock as the underlying security. The high-frequency historical data is the 1-minute data over the period from January 2, 2003 to December 31, 2012. The low-frequency historical data is the daily close prices and the option-implied volatilities are the interpolated implied volatilities of at-the-money call options with maturity being one month. The number of low-frequency observations is $n=2501$, the number of daily high-frequency observations is $m=390$, and the all the number of high-frequency prices is 975390. The in-sample period starts from January 2, 2003 to December 31, 2007, and the out-of sample period starts from January 2, 2008 to December 31, 2012. Table \ref{table_error 3 for APPLE} provides the forecasting errors of different models.

\begin{table}[!htb]
\caption{The forecasting errors of different models for APPLE stock. }
\begin{tabular}{cccccc}
\hline
  \hline
  &MAE& MSE  & AMAPE &   LL\\
  \hline
GARCH-It\^{o} &1.9735e-04 & 3.0305e-07& 0.3086 & 0.7855 \\
GARCH-It\^{o}-OI &{\bf1.9108e-04} & {\bf2.8258e-07}& {\bf   0.3053} & {\bf 0.7699} \\
GARCH-It\^{o}-IV &{\bf1.9271e-04} & {\bf2.8449e-07}&   {\bf   0.3064} & {\bf 0.7707} \\
\hline
HAR-RV &\text{\color{red}1.8432e-04} & \text{\color{red}2.5973e-07}& \text{\color{red}0.2967} & \text{\color{red}0.7264}\\
HAR-RV-OI &\text{\color{red}1.8239e-04} &\text{\color{red}2.5462e-07} &   \text{\color{red}0.2960} &  \text{\color{red}0.7136}\\
Realized GARCH(RV) & 2.8141e-04& 2.3073e-07&  0.4195 &  1.1344 \\
Realized GARCH-OI &2.8309e-04 & 3.4627e-07&  0.4068 & 1.3737 \\

\hline
GARCH+OI &3.0657e-04 & 2.7208e-07 & 0.4628 & 1.6915 \\
Realized GARCH(IV) &3.4463e-03&  4.5056e-05&  0.6243 & 7.8560 \\
\hline
IV  &2.8480e-04 &4.0924e-07 & 0.3965 & 1.4957\\
   \hline
   \hline
\end{tabular}
\label{table_error 3 for APPLE}
\end{table}

Since the addition of option-implied information, GARCH-It\^{o}-OI model and GARCH-It\^{o}-IV model has better forecasting performance than GARCH-It\^{o} model. However, they are both defeated by HAR-RV model and HAR-RV-OI model, which implies that when the high-frequency sampling interval is 1 minute, the dynamic structure of the high-frequency data in GARCH-It\^o type models is not helpful in prediction. This result is consistent with the simulation results in Subsection 5.2.1 and 5.2.2.

We also check the forecasting errors of different models during and after the subprime mortgage crisis. The results are reported in Table \ref{table_error_4 for APPLE}. We can see that HAR-RV model and HAR-RV-OI model have smaller forecasting errors than GARCH-It\^{o}-OI model and GARCH-It\^{o}-IV model after the subprime mortgage crisis, but this performance advantage is not obvious during the subprime mortgage crisis. Similar to S\&P500 index future, we also find that most of models have better MAE and MSE, but worse AMAPE and LL after subprime mortgage crisis.

\begin{table}[!htb]
\footnotesize
\caption{The forecasting errors during and after the subprime mortgage crisis for APPLE stock. }
\begin{tabular}{cccccccccc}
\hline
\hline
& \multicolumn{4}{c}{During the subprime mortgage crisis} & \multicolumn{4}{c}{After the subprime mortgage crisis}\\
  \hline
 &MAE & MSE  & AMAPE &   LL & MAE & MSE & AMAPE &   LL\\
  \hline
 GARCH-It\^{o} &4.059e-04&7.893e-07 & 0.270 & 0.569 &9.354e-05&3.127e-08& 0.314 & 0.836\\
GARCH-It\^{o}-OI& {\bf3.917e-04} & {\bf 7.299e-07} & {\bf0.268}& {\bf 0.560} & {\bf9.196e-05} & {\bf3.031e-08} & {\bf  0.311} & {\bf  0.821} \\
GARCH-It\^{o}-IV& {\bf3.956e-04}& {\bf7.355e-07}& {\bf 0.269} & {\bf 0.562} &{\bf9.215e-05}& {\bf3.044e-08} & {\bf  0.312} & {\bf 0.822} \\
\hline
HAR-RV &4.065e-04& \text{\color{red}6.742e-07}& 0.284 &   0.613&\text{\color{red}8.175e-05}& \text{\color{red}2.338e-08}&  \text{\color{red}0.299} & \text{\color{red}0.769}\\
HAR-RV-OI &3.928e-04&\text{\color{red}6.579e-07}& 0.272&  0.572&\text{\color{red}8.060e-05}&\text{\color{red}2.274e-08}&\text{\color{red}0.293} & \text{\color{red}0.688}\\
Realized GARCH(RV)&5.926e-04&  6.162e-07 & 0.395 & 1.004 &1.503e-04& 3.199e-08 & 0.435& 1.167\\
Realized GARCH-OI& 6.245e-04&  9.186e-07& 0.410 & 1.281&1.354e-04&3.400e-08 &  0.419 & 1.480 \\
\hline
GARCH+OI & 4.792e-04 &   6.067e-07 & 0.344 & 0.838 &  1.919e-04 &  5.797e-08&  0.490 &  1.832\\
Realized GARCH(IV)  &6.900e-04  & 1.049e-06 & 0.431 & 1.556  & 7.601e-03 & 1.112e-04&  0.737 & 14.941 \\
\hline
IV  & 4.438e-04&  8.796e-07 & 0.302&  0.724& 1.005e-04 &2.827e-08 &   0.378 & 1.287\\
   \hline
   \hline
\end{tabular}
\label{table_error_4 for APPLE}
\end{table}

\subsection{Sugar future}
The third security studied is Sugar future in China. As the Sugar future option was first traded on April 19, 2017, the high-frequency historical data is the 5-minute data over the period from May 2, 2017 to August 31, 2018. The low-frequency historical data is the daily close prices and the option-implied volatilities are the interpolated implied volatilities of at-the-money call options with maturity being one month. The number of high-frequency data is 14445 (321 days). The in-sample period starts from May 2, 2017 to December 31, 2017, and the out-of sample period starts from January 2, 2018 to August 31, 2018.  The forecasting errors of different models are reported in Table \ref{table_error 1 for Sugar Option}. We can see that the GARCH-It\^o-OI model has the stronger prediction power than other volatility models, which is consist with the results for S\&P500 index future.

\begin{table}[!htb]
\caption{The forecasting errors of different models for Sugar future. }
\begin{tabular}{cccccc}
\hline
  \hline
  &MAE& MSE  & AMAPE &   LL\\
  \hline
GARCH-It\^{o} &1.8752e-05 & 8.6125e-10& 0.4773 & 2.1926 \\
GARCH-It\^{o}-OI &{\bf1.6077e-05} & {\bf4.9569e-10}& {\bf  0.4451} & {\bf 1.8410} \\
GARCH-It\^{o}-IV &{\bf1.6895e-05} & {\bf5.1655e-10}&   {\bf  0.4776} & {\bf 2.1085} \\
\hline
HAR-RV &1.6633e-05 &5.0837e-10 &   0.4484 & 1.8470\\
HAR-RV-OI &1.6556e-05 & 4.9609e-10& 0.4478 & 1.8667\\
Realized GARCH(RV) & 3.0961e-05& 1.2434e-09&  0.5832 &  3.3923 \\
Realized GARCH-OI &0.0011 & 8.5602e-06& 0.7129 & 9.3393 \\

\hline
GARCH+OI &4.7113e-05 & 2.4731e-09 & 0.6559 & 5.1865 \\
Realized GARCH(IV) &2.9655e-05&  1.0538e-09&  0.5916 & 3.7273 \\
\hline
IV  &1.6108e-05 &4.9680e-10 & 0.4586 & 2.0336\\
   \hline
   \hline
\end{tabular}
\label{table_error 1 for Sugar Option}
\end{table}

\section{Conclusion}
After proposing the GARCH-It\^{o}-OI model and GARCH-It\^{o}-IV model, which are the explicit models integrating the low-frequency historical data, the high-frequency historical data and the option-implied volatility, we obtain the quasi-maximum likelihood estimators for the parameters and establish their asymptotic properties. In simulation study and empirical study, we show that the proposed GARCH-It\^{o}-OI model and  GARCH-It\^{o}-IV model has better out-of-sample forecasting performances than other models in the literature, when the high-frequency sampling interval is 5 minute. However, when the high-frequency sampling interval is 1 minute or 10 seconds, the HAR-RV model has better forecasting performance. Thus, specifying the dynamic structure of the high-frequency data as an It\^o process with time changing volatility in the GARCH-It\^o type models may not be helpful in prediction, when the high-frequency sampling interval is small. Then, how to model an explicit mixed-frequency model when the the sampling interval of the high-frequency data is small is an interesting and changeling future research direction.

The proposed GARCH-It\^{o}-OI model and GARCH-It\^{o}-IV model can be also extended in several other directions. First, the jump components of the conditional volatility, and the asymmetry between positive and negative return shocks on the conditional volatility can be added to the model, which may generate better volatility forecasts. Second, when estimating the model's parameters, a quasi-likelihood function containing realized volatility, daily log return and option-implied volatility can be constructed, which may make full use of three information sources.\footnote{We thank the anonymous referee for pointing out these two very valuable future research directions.}

\section*{Acknowledgments}
This work was partially supported by National Natural Science Foundation of China (71671106), the State Key Program of National Natural Science Foundation of China (71331006), the State Key Program in the Major Research Plan of National Natural Science Foundation of China (91546202).

\end{document}